\documentclass[conference]{IEEEtran}
\IEEEoverridecommandlockouts
\usepackage{cite}
\usepackage{amsmath,amssymb,amsfonts}
\usepackage{algorithmic}
\usepackage{graphicx}
\usepackage{textcomp}
\usepackage{booktabs}
\usepackage{xcolor}
\usepackage{soul}
\usepackage{tablefootnote}
\usepackage{threeparttable}
\usepackage[caption=false,font=footnotesize]{subfig}
\usepackage{amssymb}
\usepackage{multirow}
\usepackage{svg}
\usepackage{cleveref}

\crefname{figure}{Fig.}{Figs.}
\crefname{section}{Sec.}{Secs.}
\crefname{subsection}{Sec.}{Secs.}
\crefname{table}{Table}{Tables}
\crefname{equation}{Eq.}{Eqs.}

\def\BibTeX{{\rm B\kern-.05em{\sc i\kern-.025em b}\kern-.08em
    T\kern-.1667em\lower.7ex\hbox{E}\kern-.125emX}}
\begin{document}

\title{\huge Design Space Exploration of In-Memory Computing Implementations for Discrete Fourier Transform}

\author{
\IEEEauthorblockN{Sofia Tatidis\textsuperscript{*}, Philip Nielsen\textsuperscript{*}, Liang Liu \textsuperscript{*}, and Mojtaba Mahdavi\textsuperscript{§}}
\IEEEauthorblockA{\textsuperscript{*}Dept. of Electrical and Information Technology, Lund University, Sweden \\
\textsuperscript{§}Ericsson Research, Lund, Sweden}
}
        
\maketitle

\begin{abstract}Although memristor-based in-memory computing (IMC) has been widely investigated for brain-inspired neuromorphic workloads, systematic evaluations of its energy, latency, and signal-to-noise ratio (SNR) trade-offs across diverse system parameters remain scarce for classical digital signal processing (DSP). To address this gap, this paper introduces a comprehensive hardware-aware design framework for systematically mapping algorithmic workloads onto IMC architectures, using the discrete Fourier transform (DFT) as a case study. Tailored to the stringent performance requirements of a DFT accelerator for 5G orthogonal frequency-division multiplexing (OFDM) systems, we propose a novel mapping scheme that reduces energy consumption per computation by 53\% compared with conventional mapping techniques. Additionally, we present a comparative study of resistive random-access memory (RRAM) and ferroelectric tunnel junction (FTJ) technologies with identical numbers of programmable states but distinct conductance ranges, demonstrating the importance of device-informed architectural co-design. By bridging the gap between emerging IMC architectures and rigid DSP constraints, this work provides a pathway toward energy-efficient edge accelerators for future wireless communication systems.

\end{abstract}

\section{Introduction}
The discrete Fourier transform (DFT) is one of the most fundamental algorithms in digital signal processing and plays a central role in modern communication and multimedia systems. Importantly, it enables orthogonal frequency-division multiplexing (OFDM), a key modulation technique in wireless communication systems supporting high data throughput. OFDM systems require real-time processing of input data within a time frame determined by the bandwidth, which also defines the size of the DFT.

The fast Fourier transform (FFT), an efficient implementation of the DFT, has significantly reduced computational complexity, making real-time signal processing feasible in a wide range of embedded and high-performance systems, including OFDM. Despite algorithmic optimizations, conventional hardware implementations of DFT architectures remain constrained by the von Neumann computing paradigm, in which memory and processing units are physically separated. This separation results in frequent and energy-intensive data transfers, leading to increased latency, higher power consumption, and limited scalability. As technology scales and data-intensive workloads continue to grow, the memory wall and bandwidth limitations have become major bottlenecks, particularly for large transform sizes and high-throughput applications.

In-memory computing (IMC) has emerged as a promising solution to these challenges by performing computations directly within memory arrays, thereby significantly reducing data movement. Memristive crossbar arrays, in particular, facilitate the efficient implementation of matrix-vector multiplications, which form the computational core of DFT operations. The feasibility of IMC-based DFT/FFT implementations using memristor crossbar arrays is evaluated in \cite{Yuan_MWSCAS_2017, Zhao_NC_2023, Wen_DATE_2024}. In  \cite{Yuan_MWSCAS_2017}, a memristor-based crossbar array DFT was simulated as an accelerator for multiple-input multiple-output (MIMO) detectors. The proposed architecture is compared with conventional CMOS implementations and shown to achieve significant improvements in latency and power for 256- to 1024-point DFTs. In \cite{Zhao_NC_2023}, a memristive image reconstructor for medical applications is implemented and evaluated, demonstrating the feasibility of DFT computation in memristor-enabled IMC. The architecture used in \cite{Yuan_MWSCAS_2017, Zhao_NC_2023} is further optimized for energy efficiency and area in \cite{Wen_DATE_2024} by introducing stacked matrices, effectively reducing peripheral overhead.

The IMC-based DFT implementation is impacted, in a nontrivial way, by design parameters in algorithm mapping, crossbar architecture, and device features. The trade-offs among energy consumption, timing performance, and signal-to-noise ratio (SNR) across different system parameters have not yet been systematically investigated. To bridge this gap, in this paper we present a design space exploration framework that enables such cross-level design trade-off and optimization for mapping DFT on a IMC platform. More specifically, we 
\begin{itemize}
    \item Propose a new architecture which leverages the symmetric properties of the DFT to halve the crossbar size.%
    \item Investigate crossbar tiling strategies to mitigate the static IR drop effect and lower the required ADC resolution.
    \item Study the impact of input/coefficient bit-widths and data flow on SNR, energy consumption and processing time.
    \item Evaluate different IMC devices and analyze the impact of device imperfections at system level, using a 1024-point DFT implementation aimed for OFDM modulation.%
\end{itemize}

\section{Background} \label{sec.Background}
\subsection{Orthogonal Frequency-Division Multiplexing} \label{sec.OFDM}
\begin{figure}
\centerline{\includegraphics[width=0.9\linewidth]{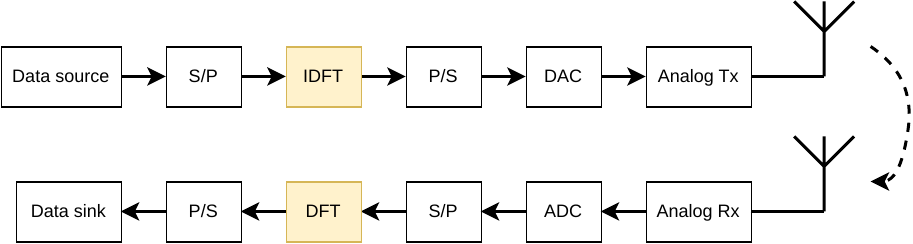}}\vspace{-7pt}
\caption{Simplified overview of an OFDM system with IDFT and DFT operations highlighted.}
\label{fig.OFDM}
\end{figure}

The principle of OFDM is to use multiple orthogonal subcarriers to modulate data for wireless transmission. In an OFDM system, as shown in \cref{fig.OFDM}, the transmitter modulates the signal by performing the inverse DFT on the frequency-encoded signal before transmission. At the receiver, the DFT is then applied to demodulate and recover the original frequency-domain signal. The sampling rate of the OFDM receiver is

\vspace{-5pt}
\begin{equation}
\label{eq.OFDM_sampling}
f_s=f_{sc}\times N_{DFT},
\end{equation}

\vspace{-5pt}where $f_{sc}$ is the subcarrier spacing and $N_{DFT}$ is the length of the DFT. The entire OFDM symbol is received after $N_{DFT}\times\frac{1}{f_s}$ seconds, at which point the receiver must be ready to process the next symbol. This, in turn, defines the throughput requirement of the DFT accelerator.

Fundamentally, the DFT computes the frequency spectrum of a discrete-time signal $\textbf{\textit{x}}=[x_0,x_1,\cdots,x_{N-1}]$ by transforming $N$ time-domain samples, $x_n$, into $N$ frequency-domain samples, $X_k$, according to \vspace{-5pt}
\begin{equation}\label{eq.FFT}
X_k=\sum_{n=0}^{N-1}x_nW^{nk}*N \vspace{-4pt}
\end{equation}
where $\textbf{\textit{X}}=[X_0,X_1,\cdots,X*{N-1}]$ denotes the DFT output, $W^{nk}_N=e^{-j2\pi nk/N}$ represents the complex-valued coefficients known as \textit{twiddle factors}, and $n,k=0,1,\cdots,N-1$. Conventional pipelined FFT implementations buffer and process the input elements according to the corresponding butterfly operation flow, whereas the IMC architectures considered in this paper require the full input vector to be available before the core computation can be performed. This necessitates careful design of the input buffering.

\subsection{In-Memory Computing}\label{sec.IMC}
In-memory computing introduces a paradigm shift by embedding computational functionality directly within memory arrays, thereby alleviating fundamental constraints of conventional architectures, particularly the \textit{von Neumann} bottleneck. Among emerging device technologies, \textit{memristors}\footnote{For conciseness, the term memristor is used throughout this work to refer broadly to the considered memristive devices.} have gained significant attention due to their ability to simultaneously store and process information in the analog domain, making them strong candidates for IMC hardware implementations~\cite{Yang_TCAS_2022}. Each memristive cell can be programmed to encode multibit data (e.g., 6-bit precision), depending on the number of stable and distinguishable conductance states it can reliably support. Technologies such as ferroelectric tunnel junctions (FTJs) and resistive random-access memory (RRAM) fall into this category and differ in material composition, switching mechanisms, scalability, and endurance characteristics~\cite{Yang_TCAS_2022}. \par

\begin{figure}
\centerline{\includegraphics[width=5.8cm]{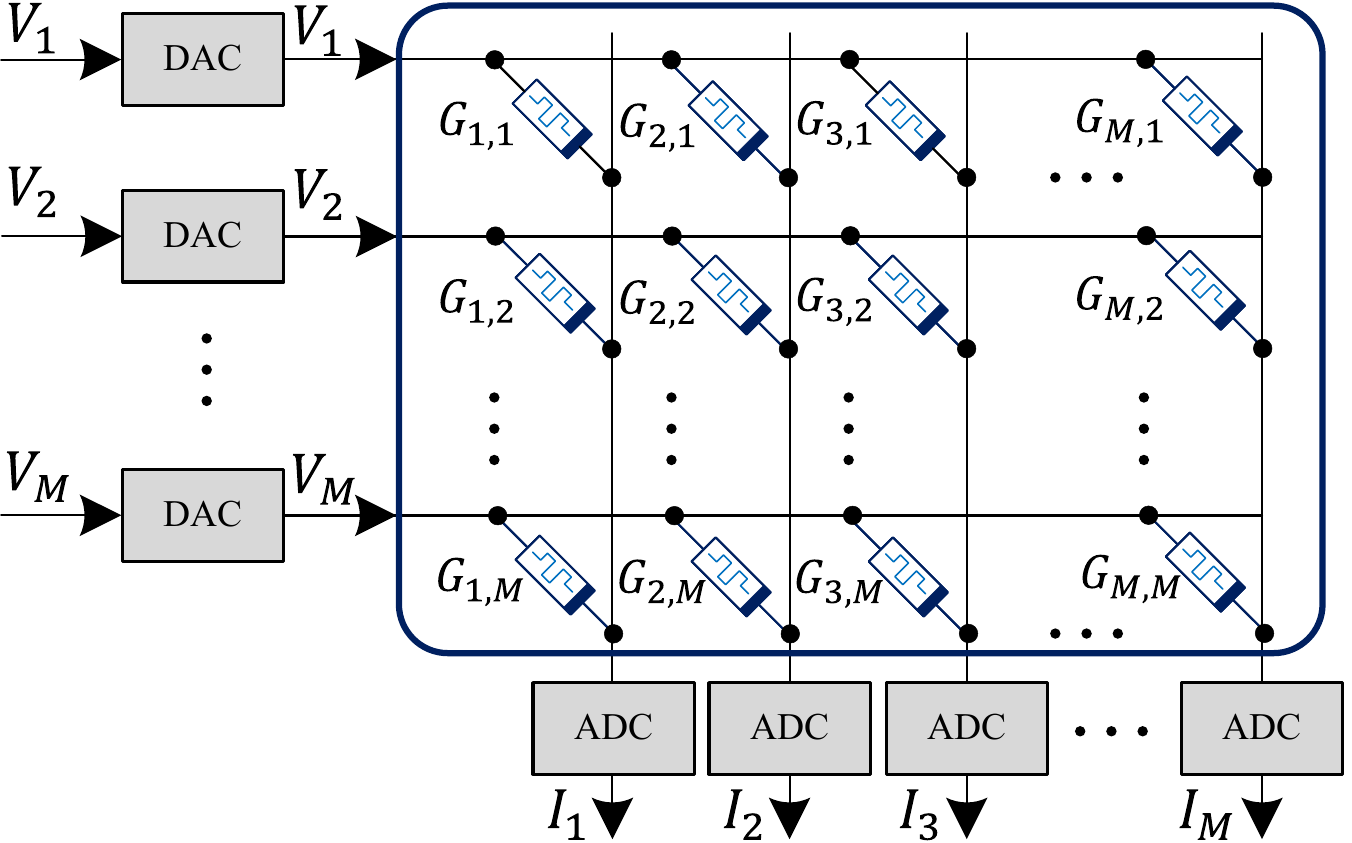}}\vspace{-7pt}
\caption{A memristor-based crossbar array of size $M\times M$.}\vspace{-7pt}
\label{Fig_Crossbar_General}
\end{figure}

\cref{Fig_Crossbar_General} illustrates an $M\hspace{-3pt}\times \hspace{-3pt}M$ \textit{crossbar array}, in which memristive elements are programmed according to the entries of an $M\hspace{-1pt}\times\hspace{-1pt} M$ conductance matrix, $\textit{\textbf{G}}$. The input vector $\textit{\textbf{V}}=[V_1,\cdots,V_M]^T$ is applied across the rows of the array. Exploiting Ohm's law for multiplication and Kirchhoff's current law for accumulation, the crossbar inherently computes the dot product between the input vector and each column of the array as \vspace{-7pt}
\begin{equation}\label{eq.Crossbar_Column}
I_i=\sum_{j=1}^{M}G_{i,j}V_j,  \hspace{+40pt} i=1,\cdots,M \vspace{-5pt}
\end{equation}
where $G_{i,j}$ denotes the element at the intersection of the $i$-th row and $j$-th column of matrix $\textbf{\textit{G}}$. Consequently, the crossbar performs the matrix-vector multiplication (MVM)\vspace{-8pt}
\begin{equation}\label{eq.Crossbar_General}
\textbf{\textit{I}}=\textbf{\textit{G}}\textbf{\textit{V}},\vspace{-7pt}
\end{equation}
where $\textbf{\textit{I}}\hspace{-3pt}=\hspace{-3pt}[\hspace{-1pt}I_1,\hspace{-1pt}\cdots\hspace{-1pt},I_M]^T\hspace{-2pt}$ represents the output current vector collected from the columns. As depicted in \cref{Fig_Crossbar_General}, interfacing the analog crossbar with digital processing units requires data conversion stages. Specifically, digital-to-analog converters (DACs) translate binary input words ($V_i$) into corresponding analog voltages, while analog-to-digital converters (ADCs) quantize the resulting column currents into digital values. These converters can be allocated per column for maximum parallelism or shared in a time-multiplexed configuration to reduce hardware overhead and power consumption.

\subsection{Key Challenges of Analog IMC}

The inherent nonidealities of analog memristive devices and IR drop in crossbar arrays degrade the signal-to-noise ratio (SNR) of IMC computations. Conductance variation and read noise can degrade the SNR by up to 20 points, while IR drop in large crossbars can result in SNR values below 0. Additionally, data conversion can dominate the overall energy consumption while remaining essential for digitally interfaced IMC. Several techniques can improve the main figures of merit (FoMs): energy per computation (EPC), time per computation (TPC), and SNR. However, improvements in one metric should not come at the expense of the others. It is therefore crucial to evaluate all three FoMs throughout the design process and consider their trade-offs to achieve realistic system-level performance.\vspace{-5pt}

\section{Design Framework}
\begin{figure}
\centering
\includegraphics[width=1\linewidth]{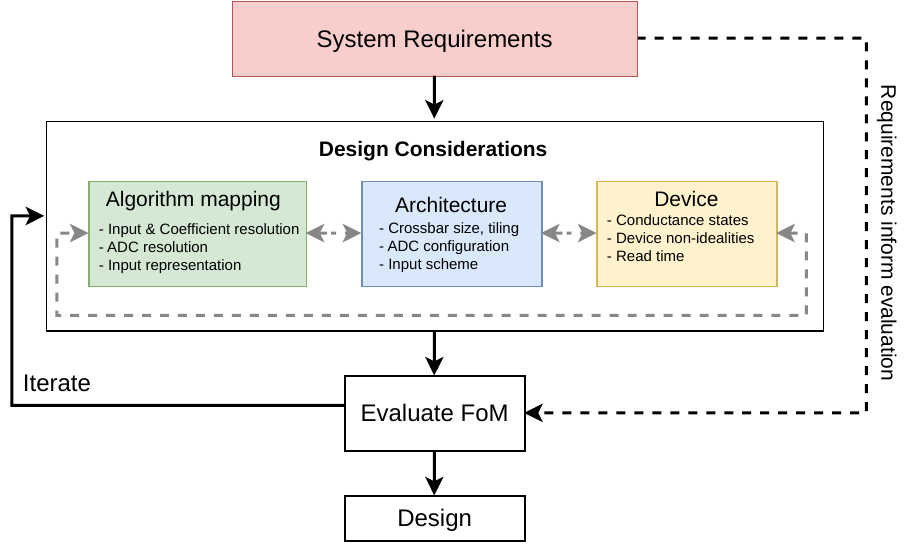}\vspace{-7pt}
\caption{The iterative cross-level design process with key considerations and evaluation points.}
\label{fig:design_framework}
\end{figure}

As outlined in \cref{fig:design_framework}, design considerations exist at several levels, ranging from algorithmic considerations such as quantization, through the physical mapping and architecture of the implementation, down to the device level. All three levels affect system performance and cannot be optimized in isolation. The main objective of this design-space exploration is to evaluate the design in terms of the FoMs against the system requirements to determine whether the target performance is met. The following sections describe these design considerations, their impact on the FoMs, and their interplay. At the end of this section, we detail the simulation process used for evaluation.

\subsection{Design Considerations}\label{sec:design_considerations}

\subsubsection{Algorithmic Mapping}
To leverage IMC for accelerating computations, coefficients are programmed into the crossbar devices at a certain resolution. Together with the input bit width, the coefficient resolution determines the level of quantization and sets the SNR ceiling of the computation, while also influencing the required ADC resolution. Inputs and coefficients can take both positive and negative values, whereas memristors can only be programmed with positive conductances. Two main methods are used to map the full value range onto the devices: offset mapping shifts all values into the positive domain, whereas differential mapping represents each coefficient using separate devices for its positive and negative components. Inputs can similarly be supplied using separate sets for positive and negative values.

\subsubsection{Architecture}
The physical mapping of the algorithm affects both energy consumption and latency. Using fewer, larger crossbars minimizes the number of costly ADCs but may increase the required ADC resolution, thereby increasing energy consumption. A large crossbar can be partitioned into tiles, where, for square tiles, the tile size $T$ denotes the number of devices per column. IMC-based architectures can support multibit input voltages; however, high input resolutions are costly because both DAC and ADC operations are energy intensive. An alternative to multibit inputs is to supply the input vector in a bit-wise manner. Under this scheme, the crossbar array processes one bit from each input word during each read cycle. After all read cycles are completed, the full DFT result is reconstructed using Shift-and-Add units placed after the ADCs. This approach requires only 1-bit DACs due to the serial nature of the inputs, resulting in notably lower hardware overhead than the parallel-input approach, at the cost of longer computation time. The ADCs can either be shared among columns to minimize chip area or individually allocated for latency-critical applications.

\subsubsection{Device}
Each memristive device is limited by the number of distinguishable conductance states that can be reliably programmed and is subject to nonidealities, which together shape the system-level performance of IMC. Read noise refers to fluctuations in the readout values and depends on the read time, read voltage, and device conductance. Longer read times can reduce read noise without necessarily becoming the dominant contributor to TPC. Conductance variation refers to differences in conductance among devices programmed to the same state, arising from material properties and the programming technique. IR drop in large crossbars is also a major contributor to SNR degradation and is influenced by the device conductance and wire resistance.

While reprogrammability and write endurance are important considerations for neuromorphic computing, they are not considered in this evaluation because the DFT coefficients remain constant. An extension of this design framework could incorporate device reprogramming for adaptive algorithms.

\subsubsection{Cross-level Interplay}

As memristors support a limited number of programmable states, careful mapping is required to represent coefficients at the desired resolution. To increase the coefficient resolution beyond that supported by a single device, each coefficient can be partitioned into smaller segments and mapped onto multiple devices, a technique referred to as coefficient slicing. However, using multiple devices per coefficient increases the required crossbar size, which can lead to non-negligible SNR degradation due to IR drop. This demonstrates how optimizing one design aspect can adversely affect another. In fact, design considerations across all levels are interdependent, highlighting the need for continuous evaluation of the FoMs throughout the design process to achieve balanced performance within the system requirements.

\subsection{Simulation Environment}
To evaluate the FoMs, a simulation framework is established using DNN+NeuroSim V1.5~\cite{Read_TCAD_2026} as the primary simulation engine. NeuroSim provides a behavioral simulator for SNR evaluation and a hardware analyzer for evaluating EPC and TPC. To address limitations in static IR drop modeling, we augment the framework with Badcrossbar~\cite{Joksas_SX_2020}. NeuroSim is also extended to support the evaluation of crossbar tiling.

The SNR of the DFT is evaluated by simulating the analog MVM. During crossbar initialization, each memory cell is assigned either a normal state or a permanent fault state based on the stuck-at-fault rate. Noise is added to the ideal conductance state of each element $G_{i,j}$ in the conductance matrix as \vspace{-5pt}
\begin{equation}\label{eq.noise}
\begin{aligned}
    G_{i,j} &= G_\text{prog}+\mathcal{N}(G_\text{prog}, \sigma^2_\text{read}), \\
    G_\text{prog} &= (G_\text{i,j,ideal}+\mathcal{N}(G_\text{i,j,ideal}, \sigma^2_\text{prog}))\times (\frac{t}{t_0})^v 
\end{aligned} \vspace{-5pt}
\end{equation} 
where $G_{i,j,ideal}$ is the ideal conductance of element $i,j$ in $\textbf{\textit{G}}$, $\sigma_\text{prog}$ and $\sigma_\text{read}$ are the standard deviations of the programming and read noise, respectively, for each state, $t$ is the retention time, $v$ is the drift coefficient, and $t_0$ is the time constant, assumed to be $1$ s. The crossbar output is calculated according to (\ref{eq.Crossbar_General}) and subsequently quantized and truncated to emulate ADC sensing.

EPC and TPC are extracted using a modified NeuroSim analyzer by isolating the SubArray module. To ensure simulation fidelity for large crossbars, the standard models are extended to include a resistive energy model for wordline/bitline ohmic losses and a first-order Elmore delay model for interconnect RC delay.

\begin{figure*}[t]
    \centering
    \subfloat[]{
        \includegraphics[width=0.31\linewidth, keepaspectratio]{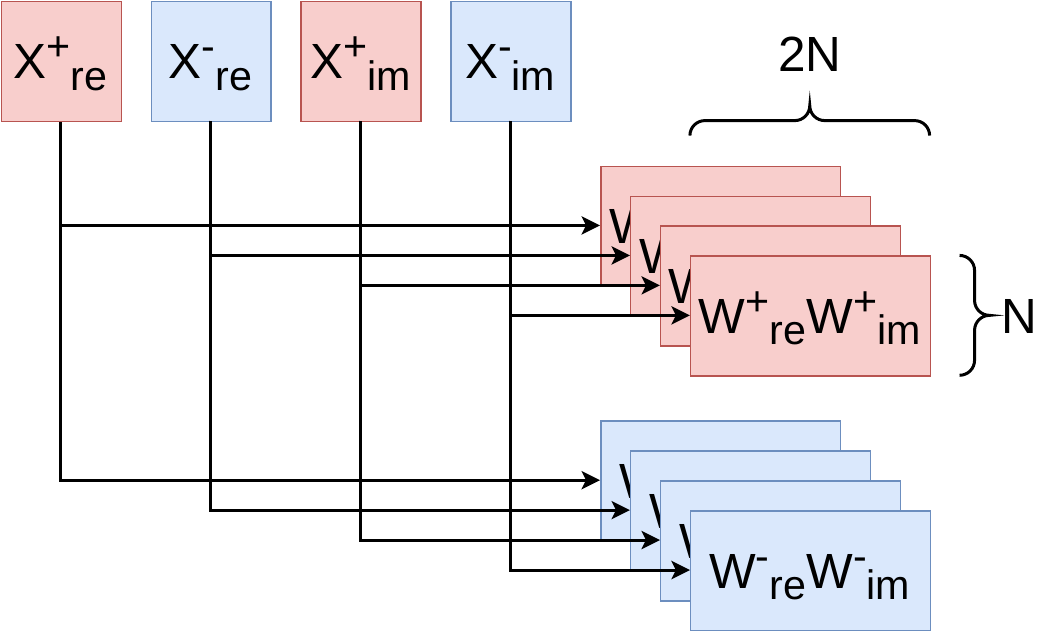}
        \label{fig:Naive}
    }
    \hfill
    \subfloat[]{
        \includegraphics[width=0.31\linewidth, keepaspectratio]{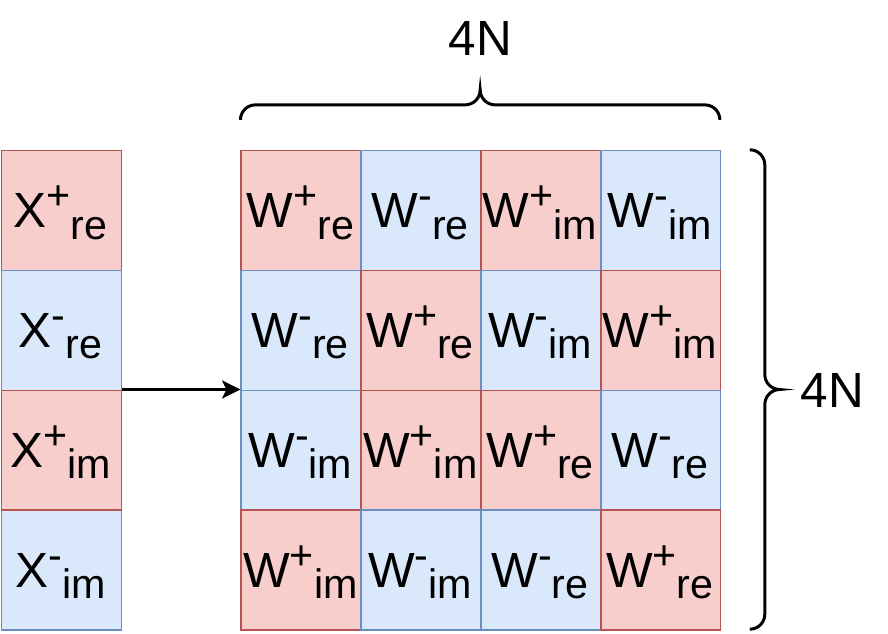}
        \label{fig:Merged}
    }
    \hfill
    \subfloat[]{
        \includegraphics[width=0.31\linewidth, keepaspectratio]{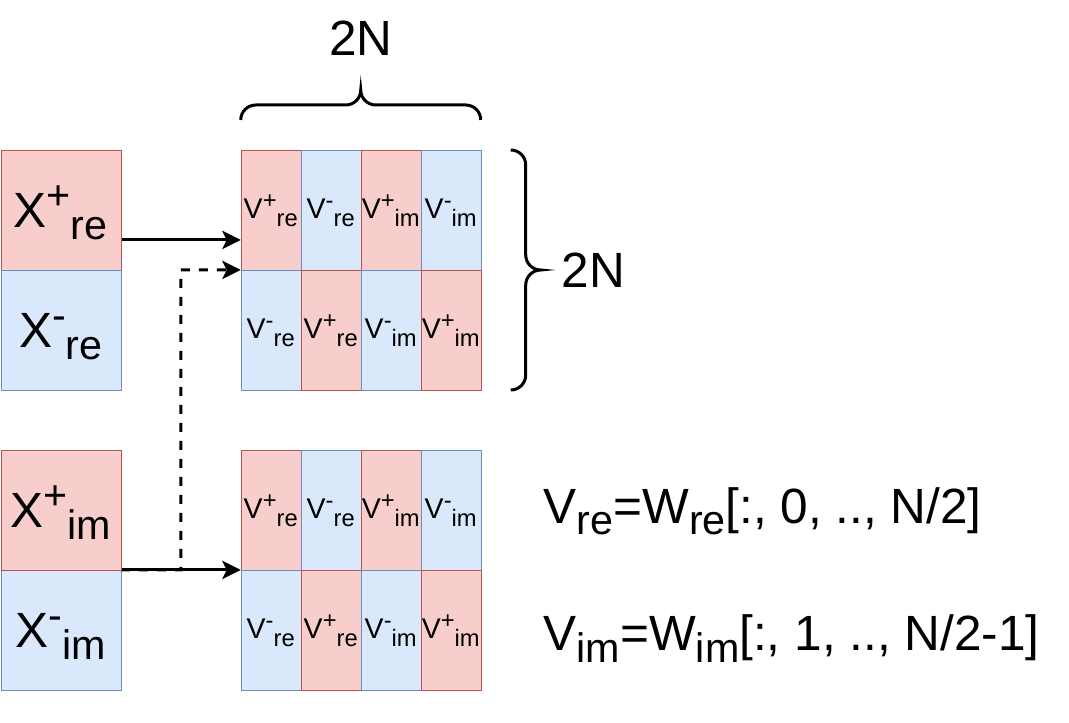}
        \label{fig:Symmetry}
    }\vspace{-4pt}
    
    \caption{Architectural progression of the IMC-based complex DFT implementations. To simplify the diagram, data conversions and digital logic are omitted. (a) Naive architecture. (b) Merged architecture. (c) Symmetry architecture.} 
    \label{fig:dft_architectures}
\end{figure*}

\section{DFT Mapping onto IMC Architectures}\label{sec.IMC_Based_DFT}
The mapping of the DFT to crossbars can be readily derived from (\ref{eq.FFT}). Three different hardware architectures for implementing the DFT using memristor crossbars are presented and compared, as illustrated in \cref{fig:dft_architectures}. All three architectures assume differential representation of the coefficients, where positive and negative values are programmed into separate devices. Although this technique can result in greater hardware overhead \cite{Wen_DATE_2024}, it has been shown to be more robust to state-independent errors than alternative approaches \cite{xiaoAccuracyAnalogNeural2022}.

\subsection{Naive Architecture}
The architecture used in \cite{Zhao_NC_2023} naively maps the DFT operation onto multiple separate crossbars (\cref{fig:Naive}). This approach requires eight $N \times 2N$ arrays to separately process the differential representations of the real and imaginary components of the input signal and twiddle factors. While conceptually straightforward, this mapping incurs a severe area and energy penalty due to the extensive use of peripheral circuitry for ADC and digital post-processing.

\subsection{Merged Architecture}
To minimize the use of costly peripherals, the computation matrices can be concatenated into a single crossbar~\cite{Wen_DATE_2024}. As shown in \cref{fig:Merged}, the entire DFT is mapped onto one large array which results in a $4N \times 4N$ crossbar array. This approach reduces the peripheral footprint compared with the Naive architecture, notably reducing the number of ADCs from $2N \times 8$ to $4N$, but suffers from substantial IR drop.

\subsection{Symmetry Architecture}
We propose a novel mapping of the DFT to address the limitations of the Naive and Merged architectures. The Symmetry architecture (\cref{fig:Symmetry}) exploits the conjugate symmetry of the DFT for real-valued inputs ($x \in \mathbb{R}^N$), for which the output satisfies $X[N-k] = X^*[k]$. Consequently, only the first half of the spectrum requires analog computation. The crossbar evaluates $N/2+1$ unique real components and $N/2-1$ unique imaginary components, effectively halving the required number of weight columns to an $N \times N$ array. The full spectrum is digitally reconstructed using the complex conjugate of the first half of the outputs by negating the imaginary components:
\vspace{-3pt}
\begin{equation}
\hspace{-6pt}X[k]\hspace{-3pt} = \hspace{-3pt}
\begin{cases} 
\hspace{-3pt}X_{Re}[k] + j \cdot X_{Im}[k], & 0 \le k \le \frac{N}{2} \\
\hspace{-3pt}X_{Re}[N-k] - j \cdot X_{Im}[N\hspace{-2pt}-\hspace{-2pt}k], & \frac{N}{2} < k \le N\hspace{-2pt}-\hspace{-2pt}1
\end{cases}
\label{eq:Symmetry_reconstruction} \vspace{-2pt}
\end{equation}

This is extended to complex inputs ($x = a + jb$) by exploiting the linearity of the DFT to independently process the real vectors $a$ and $b$ in parallel across two separate arrays (or a shared array). Let $A$ and $B$ denote the respective fully reconstructed transforms of $a$ and $b$ using (\ref{eq:Symmetry_reconstruction}). The final complex output is then digitally combined as:
\vspace{-2pt}
\begin{equation}
X_{total}[k] = \big( A_{Re}[k] - B_{Im}[k] \big) + j \cdot \big( A_{Im}[k] + B_{Re}[k] \big).
\label{eq:Symmetry_complex}
\end{equation}
This requires two $2N \times 2N$ crossbars and maintains the same number of ADCs while halving the total memristor area compared with the Merged architecture, thereby reducing the effect of IR drop.

\subsection{Peripheral Circuitry}
\subsubsection{Digital post-processing}
All three architectures require some degree of post-processing. The bit-serial input scheme requires shift-and-add operations to reconstruct the full output after all input cycles are completed. After which the crossbar outputs needs to be combined. In our investigation, this is performed digitally after the ADCs, but it could potentially be implemented using memristor-based logic as well.

\subsubsection{Input buffering}
We assume that the entire OFDM symbol is received before the input vector is transformed in a single MVM. This requires storage or buffering of the inputs until the entire symbol has been received. The time required to receive the symbol is several orders of magnitude longer than the DFT processing time. Clever system design could enable a single DFT engine to be shared among several input streams in a multiple-antenna system, but this is beyond the scope of this investigation.

\subsubsection{ADC}
The ADC is a central component of the DFT engine, as it is power-hungry, time-consuming, and contributes to quantization error. In this paper, a SAR ADC is selected and modeled in a simplified manner as specified in the user manual of the simulator \cite{Read_TCAD_2026}. The ADC resolution is derived from the formula in \cite{Wen_DATE_2024} but further optimized to account for the stacked differential coefficients, as defined by (\ref{eq.adc_resolution}):
\vspace{-5pt}

\begin{equation}
\label{eq.adc_resolution}
B_{\text{ADC}}=
\begin{cases}
 \log_2 N + b + 1, & \text{Merged architecture} \\
\log_2 N + b, & \text{otherwise}
\end{cases}
\end{equation}\vspace{-2pt}

\subsection{Device Comparison}\label{sec.FTJ_device}
Two memristor devices are compared to investigate the impact of different device properties on system performance. In \cite{Borg_NCE_2025}, a TiN/HfZrO\textsubscript{4}/W FTJ memristor \cite{Athle_AIS_2024} is identified as a promising device for analog in-memory computing, achieving 92\% classification accuracy on the MNIST dataset. The dynamic range and number of programmable states of the FTJ device were extended to 7 bits in \cite{Ozdemir_TED_2026}. A comprehensive comparison of analog resistive memories for in-memory learning found the RRAM device in \cite{Choi_NM_2018} to exhibit good performance in terms of dynamic range and endurance, two key figures of merit for online learning \cite{Xi_2021}. The FTJ device has lower conductance and read voltage, while the RRAM device has a wider dynamic range and can be programmed with a larger number of states. Key parameters used to simulate these devices are listed in \cref{tab.device_param}. \vspace{-5pt}

\begin{table}[!htp]\centering\small
\caption{Select device parameters used for simulating memristor devices.}\vspace{-8pt}
\label{tab.device_param}
\begin{tabular}{lccc}\toprule
\textbf{Device} &\textbf{Bits/device} &\textbf{G\_max/G\_min} & \textbf{Technology} \\\midrule
FTJ \cite{Borg_NCE_2025, Athle_AIS_2024, Ozdemir_TED_2026} &7 &$1.2 nS/0.12 nS$ & 22 nm\\
RRAM \cite{Choi_NM_2018} &7 &$25 \mu S/ 0.25\mu S$ & 14 nm\\
\bottomrule
\end{tabular}
\end{table}

Additionally, the devices are subject to different levels of read noise. A longer read pulse provides more robust readouts. In \cite{Lundstedt_2026}, the following relationship between read noise $r$ and read time $\tau$ is proposed for read-noise and read-time analysis:

\vspace{-5pt}
\begin{equation}
    \tau=\frac{R}{r^2V^2_{read}}(2k_BT+q_eV_{read})
\end{equation}\vspace{-2pt}

where $R$ is the average device resistance, $V_{read}$ is the read voltage, $k_B$ is the Boltzmann constant, and $q_e$ is the elementary charge. The required read pulse width for two read-noise scenarios is presented in \cref{tab:read_noise_time}.\vspace{-5pt}

\begin{table}[!htp]\centering\small
\caption{Read noise and read time scenarios}\label{tab:read_noise_time}
\begin{tabular}{lrrrrr}\toprule
\textbf{} &\multicolumn{2}{c}{\textbf{$r_A= $2\%}} &\multicolumn{2}{c}{\textbf{$r_B= $5\% }} \\\midrule
\textbf{} &\textbf{FTJ} &\textbf{RRAM} &\textbf{FTJ} &\textbf{RRAM} \\
\textbf{$\tau$} &19.9 ns &0.136 ns &2.55 ns &0.022 ns \\
\bottomrule
\end{tabular}
\end{table}\vspace{-5pt}

\section{Design Space Exploration and Performance Evaluation}\label{sec:evaluation}
Using the design considerations in \cref{sec:design_considerations}, we iteratively apply the design framework to identify a suitable design that meets the system requirements while balancing the three FoMs. The DFT for an OFDM system is used as a case study, but the design framework can be applied to different algorithms and system requirements.

\subsection{Architecture Evaluation}
\begin{figure*}[t]
    \centering

    \subfloat[]{
        \includegraphics[width=0.3\textwidth]{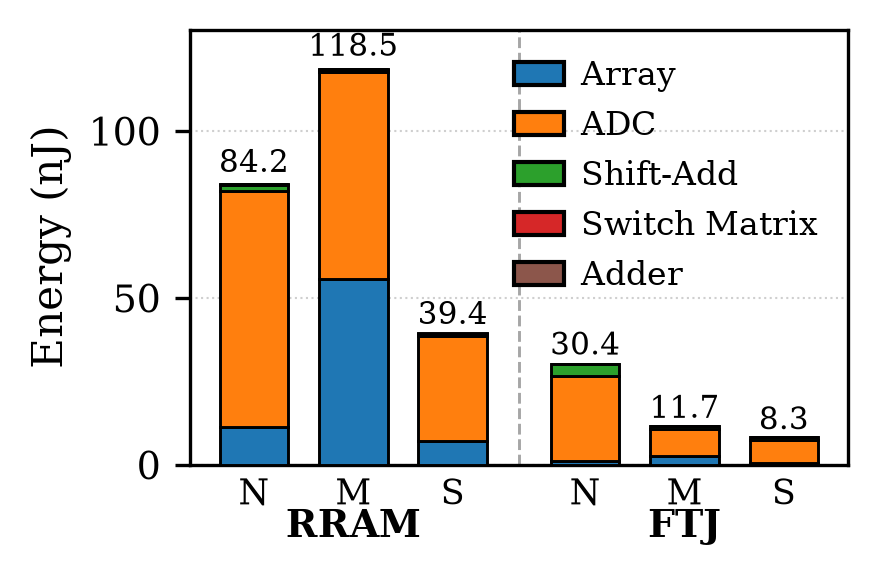}\vspace{-5pt}
        \label{fig:energy_breakdown}
        
    }
    \hfill
    \subfloat[]{
        \includegraphics[width=0.29\textwidth]{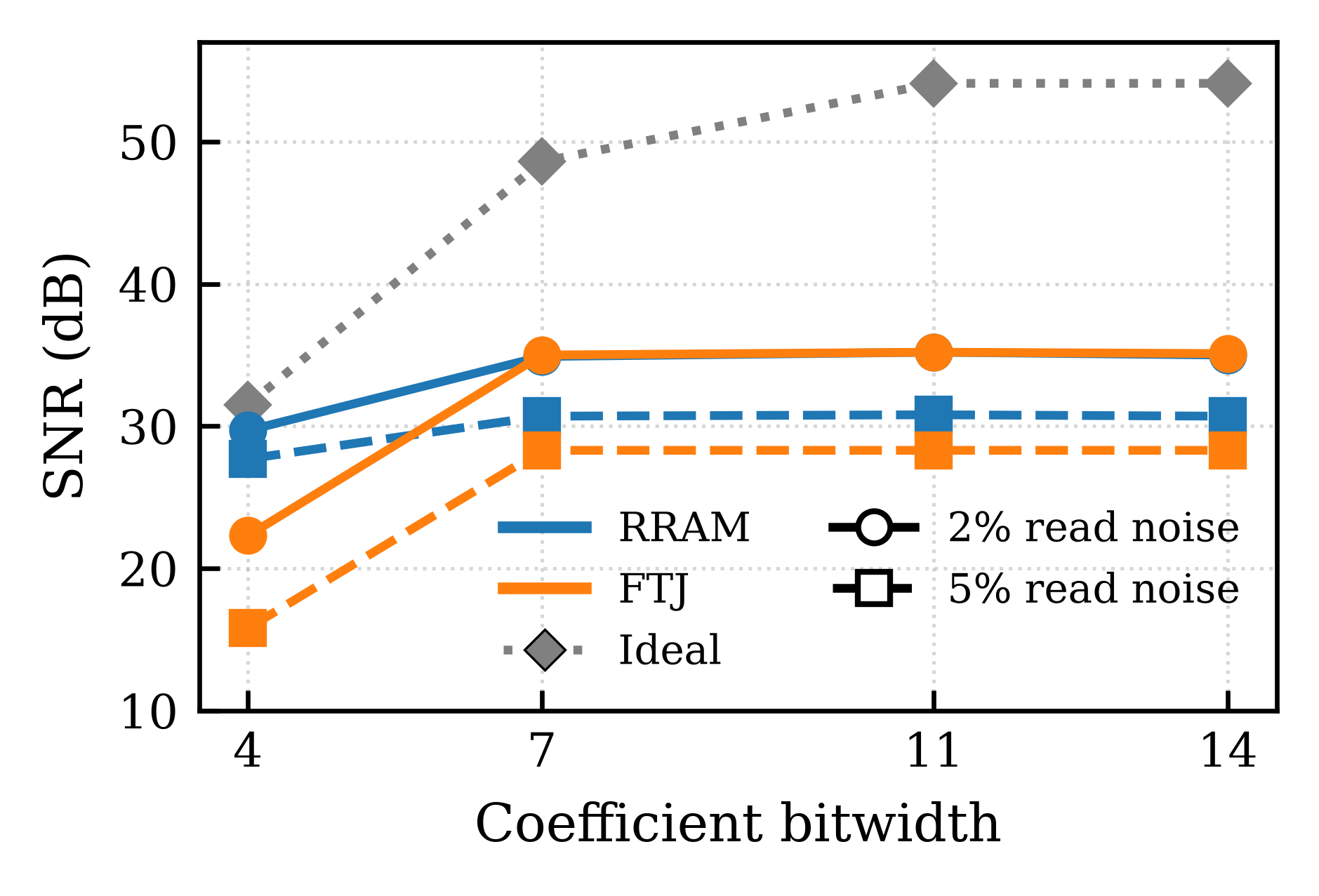}\vspace{-5pt}
        \label{fig:coeff_comp}
    }
    \hfill
    \subfloat[]{
        \includegraphics[width=0.34\textwidth]{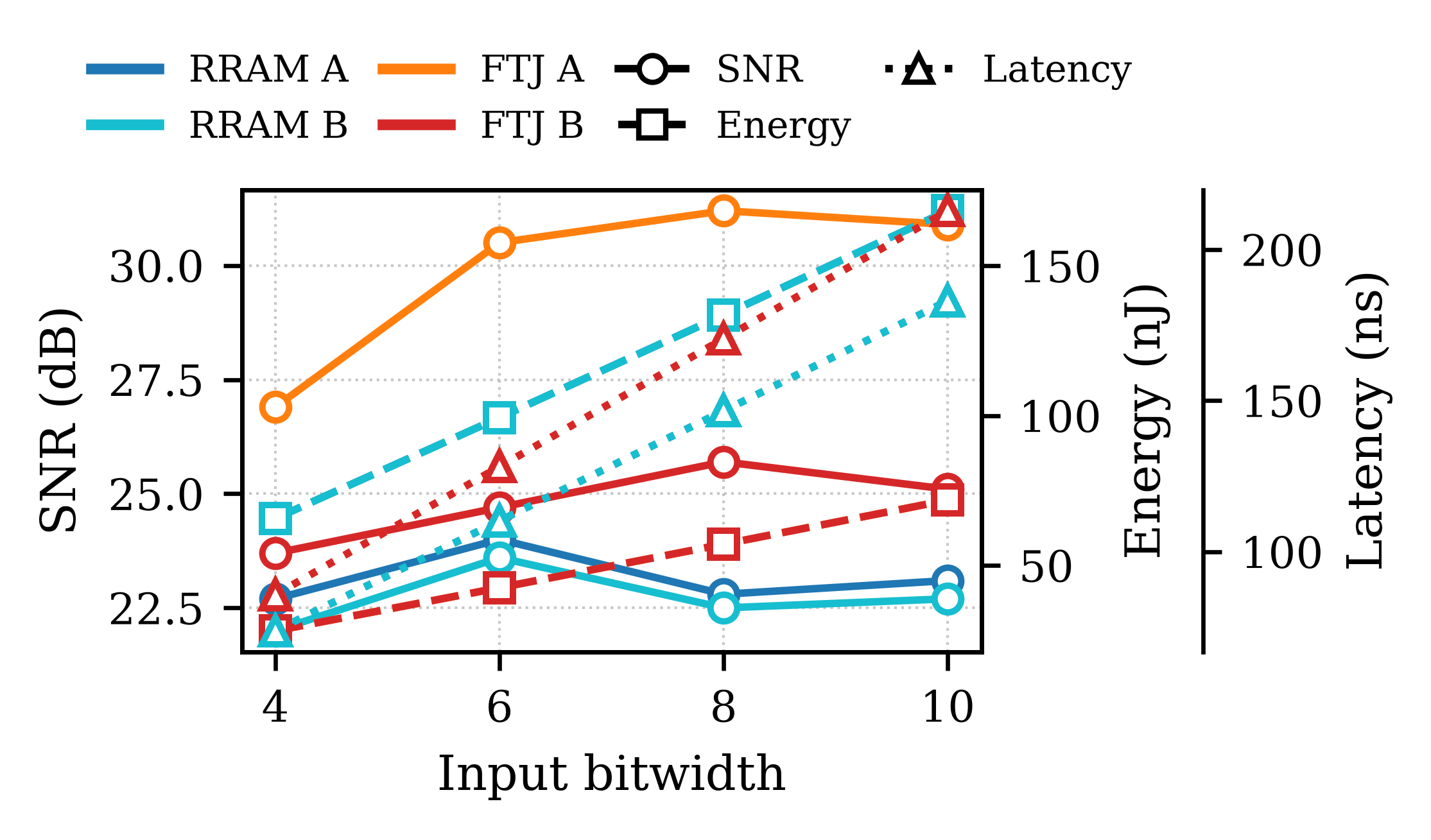}\vspace{-5pt}
        
        \label{fig:inputs_tradeoff}
    }

    \vspace{-1em}

    \subfloat[]{
        \includegraphics[width=0.30\textwidth]{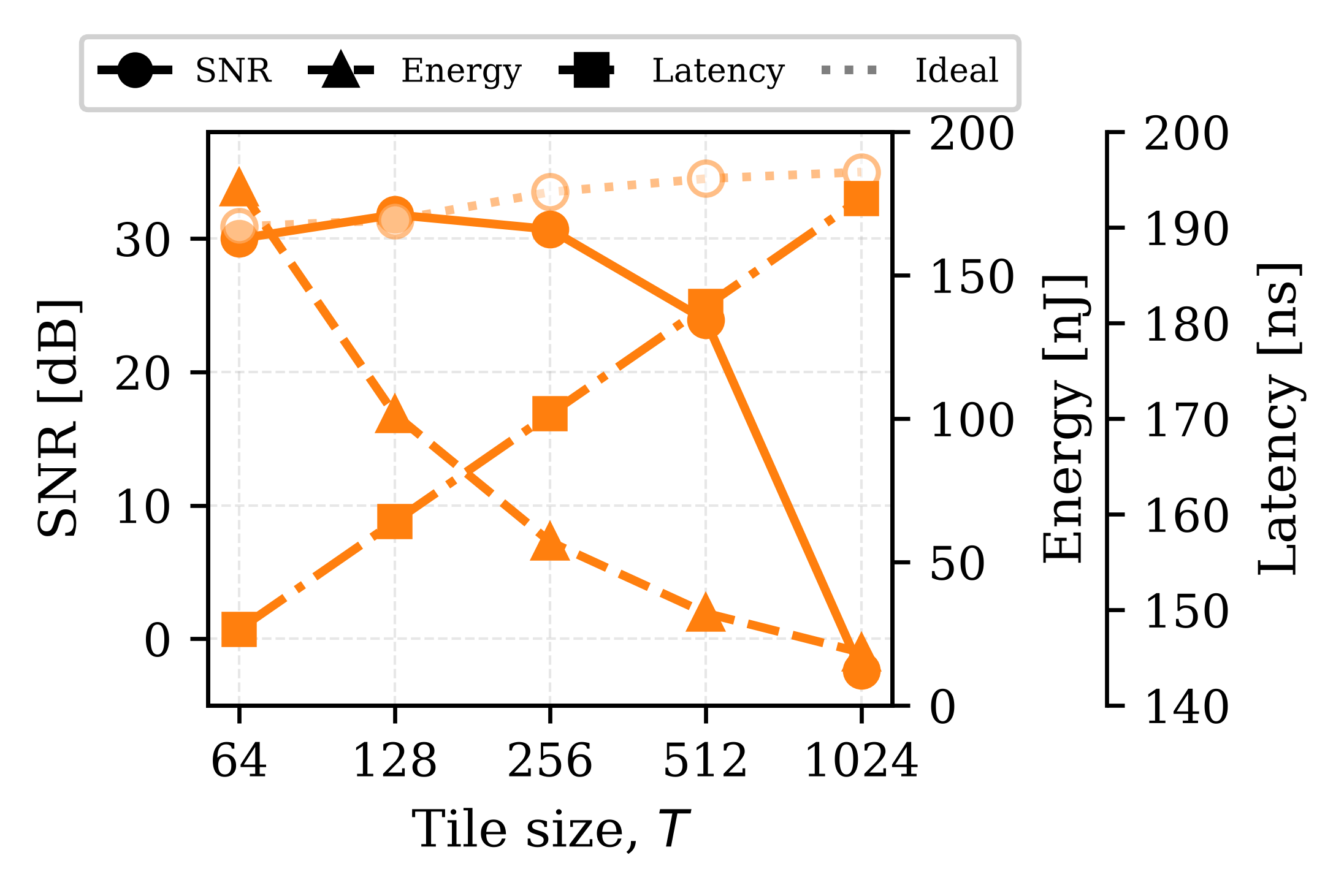}\vspace{-5pt}
        \label{fig:tile_ftj}
    }
    \hfill
    \subfloat[]{
        \includegraphics[width=0.30\textwidth]{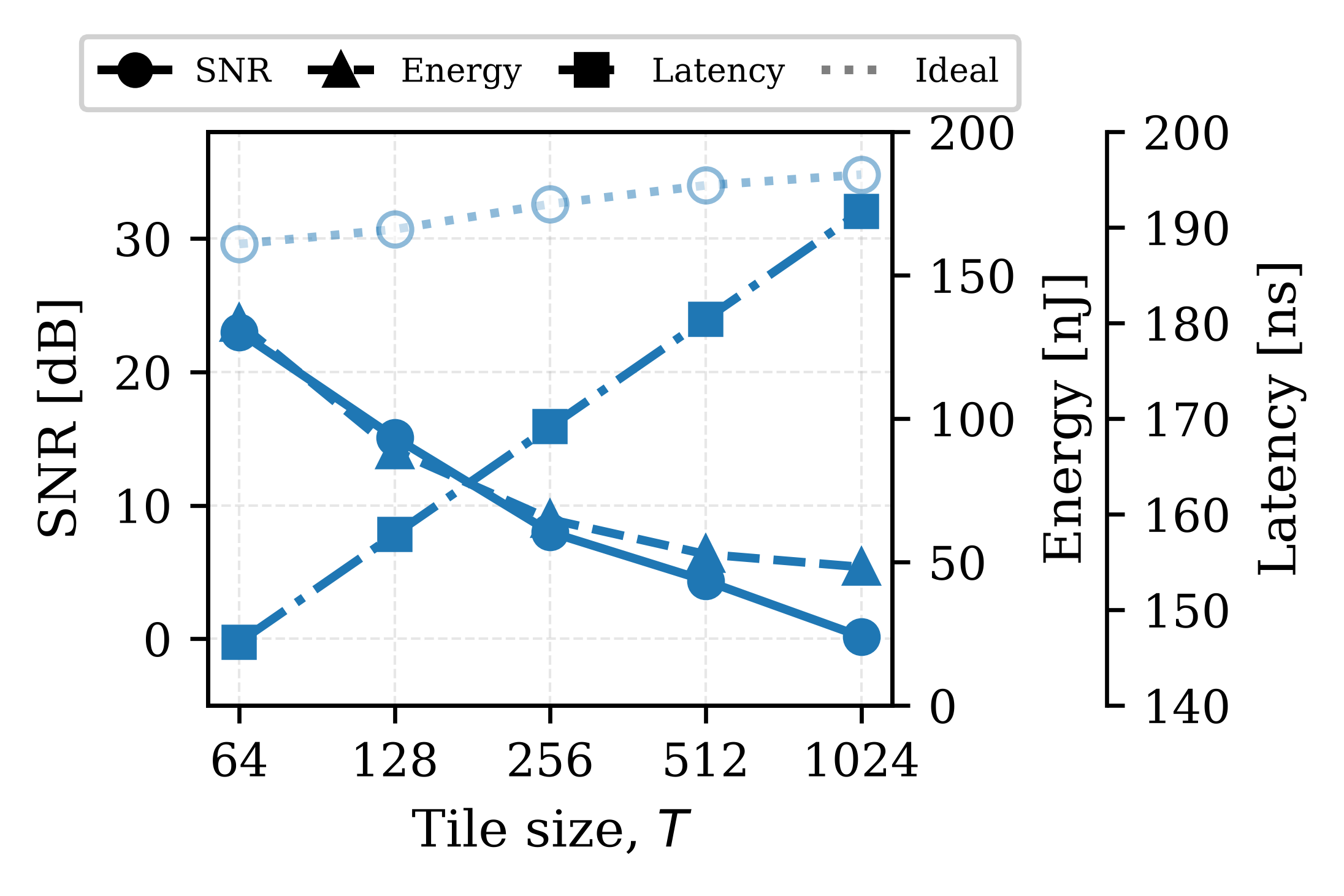}\vspace{-5pt}
        \label{fig:tile_rram}
    }
    \hfill
    \subfloat[]{
        \includegraphics[width=0.31\textwidth]{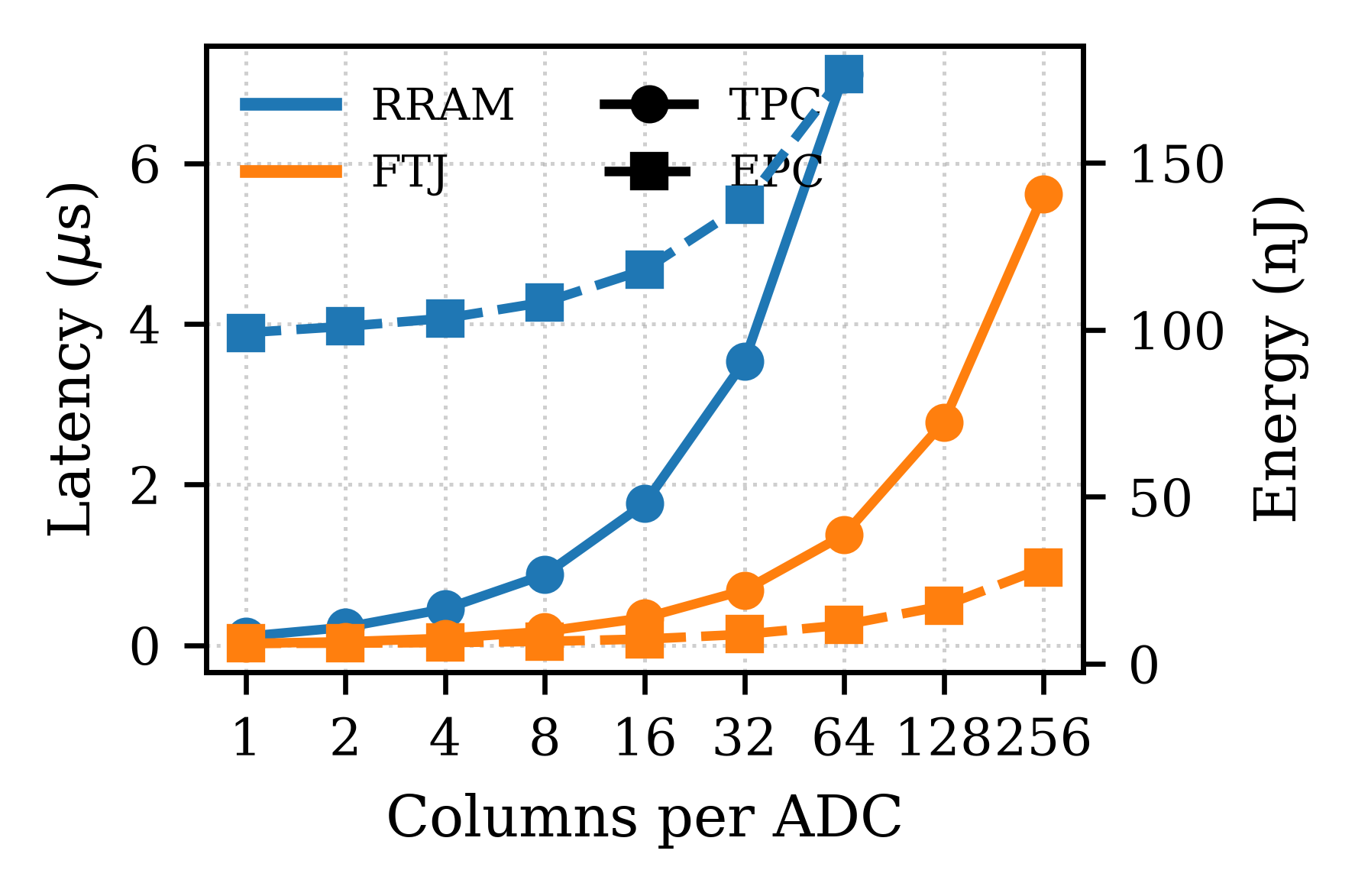}\vspace{-5pt}
        \label{fig:adc_sharing}
    }
    \vspace{-5pt}
    \caption{(a) Architecture energy breakdown, where N, M, and S denote Naive, Merged, and Symmetry, respectively. (b) Coefficient-resolution comparison. (c) Input-resolution tradeoff comparison, where A and B denote 2\% and 5\% read noise, respectively. (d)–(e) Tiling analysis for (d) the FTJ and (e) the RRAM device. (f) ADC column sharing.}
    \label{fig:ppa_res}
\end{figure*}

\vspace{-4pt}
As a case study, we consider an OFDM system with a bandwidth of 30 MHz, a subcarrier spacing of 30 kHz, and 936 subcarriers. A 1024-point DFT/IDFT is therefore appropriate for modulating the subcarriers. The three architectures outlined in \cref{sec.IMC_Based_DFT} are compared in \cref{fig:energy_breakdown}, demonstrating that the Symmetry architecture successfully eliminates the ADC-dominated energy bottlenecks by reducing the number of required ADCs. The energy breakdown differs substantially between the two devices. The RRAM device has an overall higher energy consumption, while the ADC is the dominant energy contributor for the FTJ device. Nonetheless, by exploiting conjugate symmetry to reduce the crossbar area, the Symmetry architecture achieves the lowest energy consumption with on-par latency, as shown in \cref{tab:latency_breakdown_ftj}, establishing it as the most hardware-efficient architecture across both devices. For the FTJ device, the Symmetry architecture reduces energy consumption by 29\% compared with the Merged architecture, while for the RRAM device, it achieves a 53\% reduction compared with the Naive architecture. In the subsequent analysis, the Symmetry architecture is used as the base architecture.

\subsection{Coefficient Mapping}
The number of bits used to represent the twiddle factors is a major factor in determining the SNR of an ideal memristor. Both the RRAM and FTJ devices support 7-bit programming per device, corresponding to 128 states. Four different configurations are compared, in which 4- and 7-bit coefficients are mapped to single devices, while 11- and 14-bit coefficients require two devices. For this investigation, the input resolution is fixed at 8 bits to isolate the effect of coefficient quantization to some extent. Similarly, IR drop is not simulated in order to isolate the impact of read and programming noise for the respective devices. Higher-resolution twiddle factors reduce quantization error and can potentially improve the SNR; however, as shown in \cref{fig:coeff_comp}, there is virtually no improvement in SNR beyond 7 bits per device due to device nonidealities. Using fewer states per device could potentially mitigate the impact of these nonidealities and reduce the ADC resolution requirement according to (\ref{eq.adc_resolution}). In fact, for the RRAM device, 4-bit coefficients result in only slightly lower SNR than higher resolutions and approach the performance of an ideal device. As demonstrated in the subsequent sections, the impact of IR drop is non-negligible; therefore, 7-bit devices are used in the subsequent analyses for both devices.

\begin{table}[t]
\centering
\caption{Latency breakdown for the FTJ device across architectures.}
\label{tab:latency_breakdown_ftj}
\begin{tabular}{lrrrr}
\hline
& \multicolumn{4}{c}{\textbf{Latency (ns)}} \\
\textbf{Arch.} &
\textbf{ADC} &
\textbf{Array} &
\textbf{Adder} &
\textbf{Total} \\
\hline
Naive    & 151.46 & 0.82 & 1.24 & 153.53 \\
Merged   & 159.27 & 1.41 & 1.30 & 161.99 \\
Symmetry & 151.34 & 0.94 & 1.24 & 153.53 \\
\hline
\end{tabular}
\end{table}

\subsection{Mitigating IR drop}\label{sec:ir_tile}
While the previous section considered the performance of memristor crossbars without IR drop, the SNR can be severely degraded by IR drop in large crossbars. By partitioning the crossbar into smaller tiles, acceptable SNR can be maintained. To identify a suitable tiling configuration, we limit the investigation to the 2\% read-noise scenario for simplicity. In \cref{fig:tile_ftj,fig:tile_rram}, the SNRs for different tile sizes are compared for the FTJ and RRAM devices, respectively. The results reveal two notable insights. First, SNR is severely degraded by the IR drop associated with large crossbars. For both devices, the SNR decreases by more than 20 dB when increasing the tile size from 64 to 1024. Second, the FTJ device is considerably more robust to IR drop than the RRAM device, underscoring the need for device-aware design evaluation.

For the FTJ device, the SNR varies little for tile sizes up to 256, while energy consumption decreases consistently because FTJ-based crossbars are more strongly dominated by ADC energy. This makes a tile size of 256 a suitable choice in terms of both SNR and energy. According to (\ref{eq.OFDM_sampling}), a full input vector is received every 33 $\mu$s, at which point the DFT engine must be ready to accept the next input. In terms of timing, all tile configurations complete the computation well within this limit for 8-bit inputs, providing no motivation for using smaller tiles. The trade-off between SNR and EPC differs substantially for the RRAM device, as it is far more sensitive to IR drop, experiencing a 5 dB reduction in SNR even at the smallest tile size. At the same time, the energy penalty associated with smaller tiles is less severe than for the FTJ device. Nonetheless, a tile size of 64 is required to maintain an SNR above 20 dB, which incurs an additional energy cost.

\subsection{Input Resolution and Trade-Off Discussion}
Input resolution is a key factor in determining the ideal SNR ceiling, and under the bit-wise scheme, both energy and latency scale linearly with resolution. In \cref{fig:inputs_tradeoff}, the FoMs associated with increasing the input bit width are examined for both noise scenarios. A 6-bit input provides the best SNR while maintaining low EPC and TPC. This effect is more pronounced for the FTJ device under low read noise but less noticeable at higher read-noise levels. As smaller tiles are employed for the RRAM device, the ADC resolution requirement and crossbar size are reduced, thereby lowering TPC. For both devices, the read time is negligible because latency is dominated by data conversion, as shown in \cref{tab:latency_breakdown_ftj}. This favors the longer read times to achieve lower read noise.

As observed in \cref{sec:ir_tile}, the TPC satisfies the system timing requirement with considerable margin. Because the TPC is several orders of magnitude shorter than the available processing interval, resource sharing can be employed to reduce hardware overhead. As the ADC is typically physically larger than the crossbar itself, floorplanning could benefit from sharing ADCs among crossbar columns. In \cref{fig:adc_sharing}, the number of columns sharing a single ADC is investigated, showing that a single ADC can be shared across an entire tile while satisfying the timing requirement, albeit with an energy penalty. Moreover, the available timing margin could potentially be leveraged in a MIMO system by sharing the DFT engine among multiple antennas. However, as the primary motivation for IMC is to limit data movement, the implications of such resource sharing, together with the required input buffering, must be considered before practical implementation.

\section{Conclusion}\label{sec.Conclusion}
The presented analysis demonstrates that practical IMC-based DSP design requires cross-level design-space exploration, as device characteristics and architectural choices introduce coupled trade-offs among energy, latency, and SNR. The evaluated DFT case study further demonstrates the feasibility of IMC for future wireless communication systems. More broadly, the proposed framework enables systematic identification of system-level bottlenecks and their underlying device- and architecture-level causes, providing guidance for both IMC architecture design and device-level research targeting specific DSP applications.

\bibliographystyle{IEEEtran}
\bibliography{IEEEabrv,References}

\end{document}